\documentclass{aa} 
\usepackage{float} 
\usepackage{graphicx}  
\usepackage{txfonts}  
\usepackage{natbib}
\bibpunct{(}{)}{;}{a}{}{,}
\usepackage{color}
\usepackage[normalem]{ulem}

\begin{document}  

 
\title{Accretion of planetary matter and the lithium problem in the 16 Cygni stellar system }
  
  \author{ Morgan Deal\inst{1,3}, Olivier Richard\inst{1}, \and Sylvie Vauclair\inst{2,3}}
\institute{Laboratoire Univers et Particules de Montpellier (LUPM), UMR 5299, Universit\'e de Montpellier, CNRS, Place Eug\`ene Bataillon 34095 Montpellier Cedex 5 FRANCE\     
           \and
           Universit\'e de Toulouse, UPS-OMP, IRAP, France
           \and
           CNRS, IRAP, 14 avenue Edouard Belin, 31400 Toulouse, France\\
            \email{morgan.deal@umontpellier.fr} 
           }
           
\date{\today}

\abstract
{The 16 Cygni system is composed of two solar analogues with similar masses and ages. A red dwarf is in orbit around 16 Cygni A, and 16 Cygni B
hosts a giant planet. The abundances of heavy elements are similar in
the two stars, but lithium is much more depleted in 16 Cygni B than in 16
Cygni A, by a factor of at least 4.7.}
{The interest of studying the 16 Cygni system is that the two star have the same age and the same initial composition. The differences currently observed must be due to their different evolution, related to the fact that one of them hosts a planet
while the other does not.}
{We computed models of the two stars that precisely fit the observed seismic frequencies. We used the Toulouse Geneva Evolution Code (TGEC), which includes complete atomic diffusion (including radiative accelerations). We compared the predicted surface abundances with the spectroscopic observations and confirm that another mixing process is needed. We then included the effect of accretion-induced fingering convection.}
{The accretion of planetary matter does not change the metal abundances but leads to lithium destruction, which depends upon the accreted mass. A fraction of the Earth's mass is enough to explain the lithium surface abundances of 16 Cygni B. We also checked the beryllium abundances.}
{In the case of accretion of heavy matter onto stellar surfaces, the accreted heavy elements do not remain in the outer convective zones, but are mixed downwards by fingering convection induced by the unstable $\mu$-gradient. Depending on the accreted mass, this mixing process may transport lithium down to its nuclear destruction layers and lead to an extra lithium depletion at the surface. A fraction of the Earth's mass is enough to explain a lithium ratio of 4.7 in the 16 Cygni system. In this case beryllium is not destroyed. Such a process may be frequent in planet-hosting stars and should be studied in other cases in the future.}

  
\titlerunning{Lithium destruction in the exoplanet host star 16 Cyg B}
  
\authorrunning{Deal et al.}  

\maketitle 

\section{Introduction}  

The bright solar analogues 16 Cygni A (HD 186408, HR 7503) and 16 Cygni B (HD 186427, HR 7504) for many reasons represent a very interesting stellar system. While a red dwarf, 16 Cygni C, is in orbit around the first component 16 Cygni A \citep{turner01,patience02}, the second component, 16 Cygni B, hosts a giant Jovian planet with minimum mass of 1.5 $M_{jup}$ located on an eccentric orbit (e=0.63), with an orbital period of 800.8 days \citep{cochran97}. The two main stars are separated widely enough, with an orbital period longer than 18,000 years \citep{hauser99}, to be studied in the same way as two isolated stars, with no common dynamical effects. 

This situation allows for precise differential studies between a planet-hosting star and a non-planet-hosting star with similar birth conditions. The red dwarf around 16 Cygni A may be the reason why no accretion disk has developed around it, whereas a planetary disk remained around 16 Cygni B, including the observed giant planet, and probably smaller as yet unobserved bodies.

The two main stars of the 16 Cygni system have been studied in many ways, using spectroscopy, interferometry, and asteroseismology. The abundances of the heavy elements in these two stars are very similar. Although a very small difference has been claimed by \citealt{ramirez11} and \citealt{tuccimaia14}, \cite{schuler11} found them to be indistinguishable. On the other hand, the surface lithium abundance of 16 Cygni B is lower than that of 16 Cygni A by at least a factor 4.7 \citep{king97}. These observations lead to several open questions, which remain to be answered for a better understanding of these stars. 

The interest of this study is that these stars have the same birth site and the same age, with masses of the same order, so that their past evolution is similar for most aspects. The observed differences between them must basically be due to the planetary disk around B. For this reason, the detailed study of this stellar system helps understanding the differences between stars with and without planetary disks.

The present paper is motivated by two considerations. The first one is that none of the previous modelling of these two stars
considered atomic diffusion including the radiative acceleration on each element. Most stellar evolution codes currently include the atomic diffusion of helium (without radiative accelerations), some of them also include it for heavy elements, but very few consider the radiative accelerations.

The second consideration refers to the consequences of the accretion of heavy matter onto stars, which may occur when the star has a planetary disk. Many studies and past publications assumed that the accreted matter remains inside the outer stellar convective zone, so that accretion can lead to an increase of the heavy element abundances, as well as to an increase of the lithium abundance. This assumption is not valid, as shown in detail by \citet{vauclair04}, \citet{garaud11}, \citet{theado12}, and \citet{deal13}. When heavy matter falls onto the star, it creates an inverse gradient of molecular weight, which leads to a double-diffusive instability now called fingering (or thermohaline) convection. The heavy elements are mixed downwards until the mean molecular weight gradient becomes nearly flat. In most cases, no signature of the accreted heavy elements remains at the surface.  Meanwhile, as computed in detail by \citet{theado12bis}, the induced mixing may lead to an extra lithium depletion in the star. As a consequence, the accretion of heavy matter cannot lead to any increase of lithium at the surface of a star, but may conversely lead to a decrease of its observed abundance.

The observational results on the two main stars 16 Cygni A and B are presented in Sect. 2 together with a discussion of the still-unsolved questions. In Sect. 3 we present models of 16 Cygni A and B that fit the observed seismic frequencies. We compare the parameters of these models with previously published ones. The surface abundances of heavy elements and lithium that are obtained after diffusion in these models are discussed in Sect. 4 . Finally, in Sect. 5 we show that the accretion of metal rich planetary matter at the beginning of the main sequence on 16 Cygni B may explain the lithium difference between the two stars. A summary and discussion of all these results are given in Sect. 6.

\section{Observational constraints}  

The position in the sky of the 16 Cygni system allowed seismic observations with the \textit{Kepler} satellite. More than 40 modes of degree l=0, 1, 2, and 3 could be detected. Analyses with the Asteroseismic Modelling Portal (AMP) and comparisons with other seismic studies led to precise values of the masses, radii, and ages of the two stars \citep{metcalfe12}. Moreover, the seismic observations allowed measurements of their stellar rotation periods \citep{davies15}.

The effective temperatures and gravities of the two stars were derived from spectroscopic observations (e.g. \citealt{ramirez11,schuler11,tuccimaia14}). We must insist, however, that asteroseismology leads to a log $g$ value with a much better precision than spectroscopy. As we show in Sect. 3, all the models that closely fit the observed seismic frequencies, even if they have different masses, radii, helium abundances, ages, and luminosities, have the same log $g$ value with a precision of 0.01.
The determinations of the bolometric magnitudes \citep{torres10} and Hipparcos parallaxes \citep{vanleeuwen07} lead to precise values of the luminosities of the two stars. They are also bright enough for their radii to be determined by interferometric techniques \citep{white13}. The results are given in Table \ref{table:1}.

\begin{table}
\caption{Properties of 16 Cygni A and B from the literature} 
\label{table:1}
\centering
\begin{tabular}{c c c}
\hline  
 & 16 Cygni A & 16 Cygni B   \\ 
\hline\hline
             
$T_{\rm eff}$(K) & $5825\pm 50$\tablefootmark{a} & $5750\pm 50$\tablefootmark{a}  \\
          & $5813\pm 18$\tablefootmark{b} & $5749\pm 17$\tablefootmark{b}  \\
          
          & $5796\pm 34$\tablefootmark{c} & $5753\pm 30$\tablefootmark{c}  \\
          & $5839\pm 42$\tablefootmark{d} & $5809\pm 39$\tablefootmark{d}  \\
\smallskip
          & $5830\pm 7$\tablefootmark{f} & $5751\pm 6$\tablefootmark{f}  \\
       
log $g$ & $4.33\pm 0.07$\tablefootmark{a} & $4.34\pm 0.07$\tablefootmark{a}  \\
       & $4.282\pm 0.017$\tablefootmark{b} & $4.328\pm 0.017$\tablefootmark{b}  \\
       & $4.38\pm 0.12$\tablefootmark{c} & $4.40\pm 0.12$\tablefootmark{c}  \\
       \smallskip
       & $4.30\pm 0.02$\tablefootmark{f} & $4.35\pm 0.02$\tablefootmark{f}  \\ 

[Fe/H] & $0.096\pm 0.026$\tablefootmark{a} & $0.052\pm 0.021$\tablefootmark{a}  \\
       & $0.104\pm 0.012$\tablefootmark{b} & $0.061\pm 0.011$\tablefootmark{b}  \\
       & $0.07\pm 0.05$\tablefootmark{c} & $0.05\pm 0.05$\tablefootmark{c}  \\
\smallskip      
       & $0.101\pm 0.008$\tablefootmark{f} & $0.054\pm 0.008$\tablefootmark{f}\\
\smallskip        
A(Li) & $1.27\pm0.05$\tablefootmark{i} & $\leq0.6$\tablefootmark{i}  \\ 
\smallskip        
A(Be) & $0.99\pm0.08$\tablefootmark{j} & $1.06\pm0.08$\tablefootmark{j}  \\
\hline\hline    
Mass ($M_{\odot}$) & $1.05\pm 0.02$\tablefootmark{b} & $1.00\pm 0.01$\tablefootmark{b}  \\        
                  & $1.07\pm 0.05$\tablefootmark{d} & $1.05\pm 0.04$\tablefootmark{d}  \\
\smallskip        
                  & $1.11\pm 0.02$\tablefootmark{g} & $1.07\pm 0.02$\tablefootmark{g}  \\

Radius ($R_{\odot}$) & $1.218\pm 0.012$\tablefootmark{d} & $1.098\pm 0.010$\tablefootmark{d}  \\        
                    & $1.22\pm 0.02$\tablefootmark{e} & $1.12\pm 0.02$\tablefootmark{e}  \\
\smallskip
                    & $1.243\pm 0.008$\tablefootmark{g} & $1.127\pm 0.007$\tablefootmark{g}  \\
\smallskip
Luminosity ($L_{\odot}$) & $1.56\pm 0.05$\tablefootmark{g} & $1.27\pm 0.04$\tablefootmark{g}  \\        

Age (Gyr) & $7.15_{-1.03}^{+0.04}$\tablefootmark{b} & $7.26_{-0.33}^{+0.69}$\tablefootmark{b}  \\ 
\smallskip      
          & $6.9\pm 0.3$\tablefootmark{g} & $6.7\pm 0.4$\tablefootmark{g}  \\
\smallskip
$Z_{i}$ & $0.024\pm 0.002$\tablefootmark{g} & $0.023\pm 0.002$\tablefootmark{g}  \\        
\smallskip
$Y_{i}$ & $0.25\pm 0.01$\tablefootmark{g} & $0.25\pm 0.01$\tablefootmark{g}  \\        
\smallskip
$v~\rm{sin}$ $i$ (km.s$^{-1}$) & $2.23\pm 0.07$\tablefootmark{h} & $1.27\pm 0.04$\tablefootmark{h}  \\        
\smallskip
$P_{rot}$ (days) & $23.8_{-1.8}^{+1.5}$\tablefootmark{h} & $23.2_{-3.2}^{+11.5}$\tablefootmark{h}  \\     
\smallskip
Planet detected & no & yes\tablefootmark{k}  \\   
\hline
\end{tabular}
\tablefoot{\tablefoottext{a}{\cite{ramirez09}}; \tablefoottext{b}{\cite{ramirez11}}; \tablefoottext{c}{\cite{schuler11}}; \tablefoottext{d}{\cite{white13}, seismic determination}; \tablefoottext{e}{\cite{white13}, interferometric determination}; \tablefoottext{f}{\cite{tuccimaia14}}; \tablefoottext{g}{\cite{metcalfe12}}; \tablefoottext{h}{\cite{davies15}}; \tablefoottext{i}{\cite{king97}}; \tablefoottext{j}{\cite{deliyannis00}}; \tablefoottext{k}{\cite{cochran97}}\\
}
\end{table}

Detailed determinations of their element abundances have been given by several authors. From spectroscopy with high signal-to-noise ratio of the 10 m Keck 1 telescope and HIRES echelle spectrograph, \citet{schuler11} determined the abundance of 15 heavy elements in both stars and found them indistinguishable.  On the other hand, \citet{ramirez11} and \citet{tuccimaia14} claimed that 16 Cygni A is slightly more metal rich than 16 Cygni B based
on spectra with high resolution and high signal-to-noise ratio
that were obtained with the R.G. Hull coude spectrograph on the 2.7m Harlan Smith telescope at Mc Donald Observatory. 

A more striking difference between the two stars, confirmed by all spectroscopic observations, is that the star hosting a planet, 16 Cygni B, has an abundance of lithium at least four times lower than the star without a planet (\citealt{friel93,king97}). While both stars are lithium depleted compared to F stars and to the meteoritic value, detailed measurements show that 16 Cygni A is slightly less depleted than the Sun (log $N(Li) = 1.27$ compared to 1.05 for the Sun), whereas 16 Cygni B is more depleted (log $N(Li) < 0.60$). On the other hand, \citet{deliyannis00} found that the Be and B abundances are the same in the two stars within the limits of the uncertainties.

\section{Asteroseismic studies and stellar models including radiative accelerations}  

\subsection{Stellar models}

We used the Toulouse Geneva Evolution Code (TGEC) to compute stellar models that fit the seismic observations of 16 Cygni A and B. This code performs complete computations of atomic diffusion, including radiative accelerations, for 21 species, namely 12 elements and their main isotopes: H, $^{3}$He, $^{4}$He, $^{6}$Li, $^{7}$Li, $^{9}$Be, $^{10}$B, $^{12}$C, $^{13}$C, $^{14}$N, $^{15}$N, $^{16}$O, $^{17}$O,$^{18}$O, $^{20}$Ne, $^{22}$Ne, $^{24}$Mg, $^{25}$Mg, $^{26}$Mg, $^{40}$Ca and $^{56}$Fe \citep{theado12}. The diffusion coefficients used in the code are those derived by \citet{paquette86}. 

The Rosseland opacities are recalculated inside the model, at each time step and at every mesh point, using OPCD v3.3 and data from \citet{seaton05}, to take into account the local chemical composition. In this way, the stellar structure is consistently computed all along the evolutionary tracks, as well as the individual radiative accelerations of C, N, O, Ne, Mg, Ca, and Fe. This is done by using the improved semi-analytical prescription proposed by \citet{alecian04}. A more detailed discussion of these computations is given in \citet{dealprep} (in prep).

The equation of state used in the code is the OPAL2001 equation \citep{rogers02}.
The nuclear reaction rates are from the NACRE compilation \citep{angulo99}. The mixing length formalism is used for the convective zones with a mixing length parameter of 1.8, as needed to reproduce solar models.

\begin{figure}
\begin{center}
\includegraphics[width=0.5\textwidth]{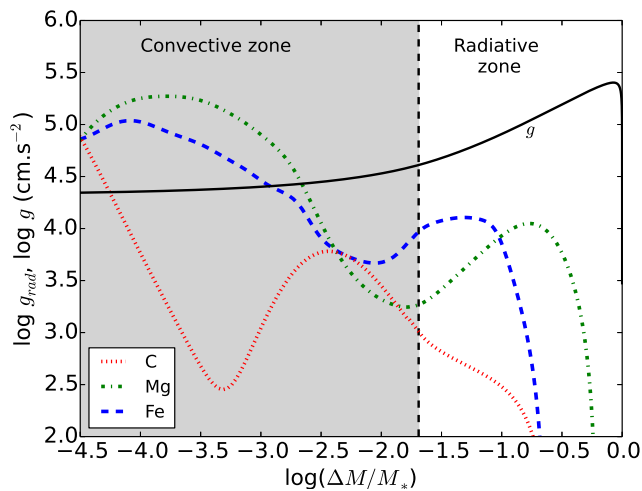}
\caption{Profiles of the radiative accelerations of three elements Fe, Mg, and C as a function of the mass fraction in 16 Cygni A. }
\label{grad}
\end{center}
\end{figure}

As an example, Fig. \ref{grad} displays the $g_{rad}$ profiles for C, Mg, and Fe compared to gravity for one of the computed models of 16 Cygni A. This model best fits the asteroseismic observations, as we discuss in Sect. 3.2. The dashed vertical line represents the position of the surface convective zone. Below the convective zone $g_{rad}$ is at least 1/3 smaller than $g$, so that the effect of the radiative accelerations is small and generally negligible inside these stars.

\subsection{Best models from asteroseismic fits}

\begin{figure*}
\begin{center}
\includegraphics[width=0.495\textwidth]{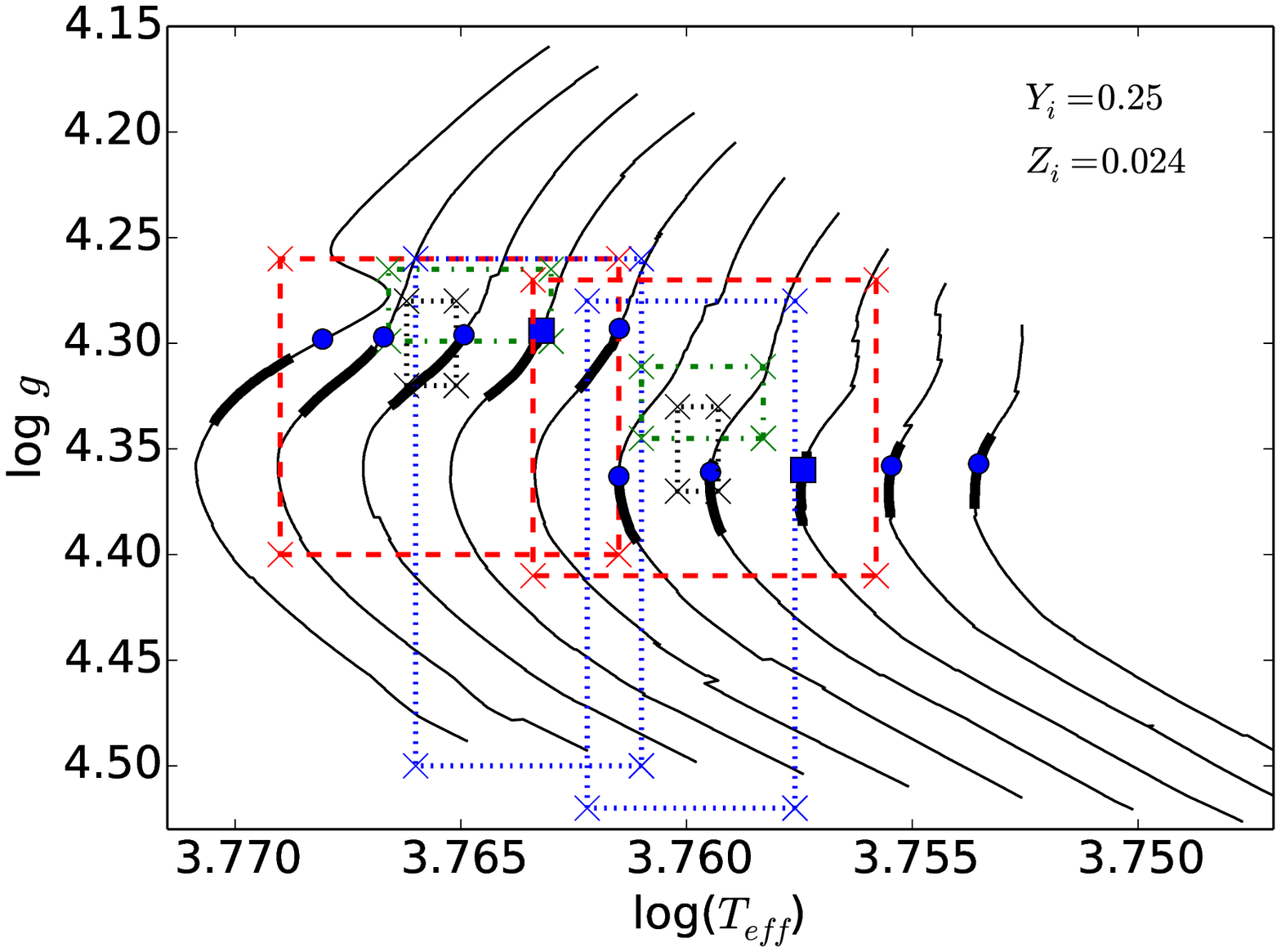}
\includegraphics[width=0.495\textwidth]{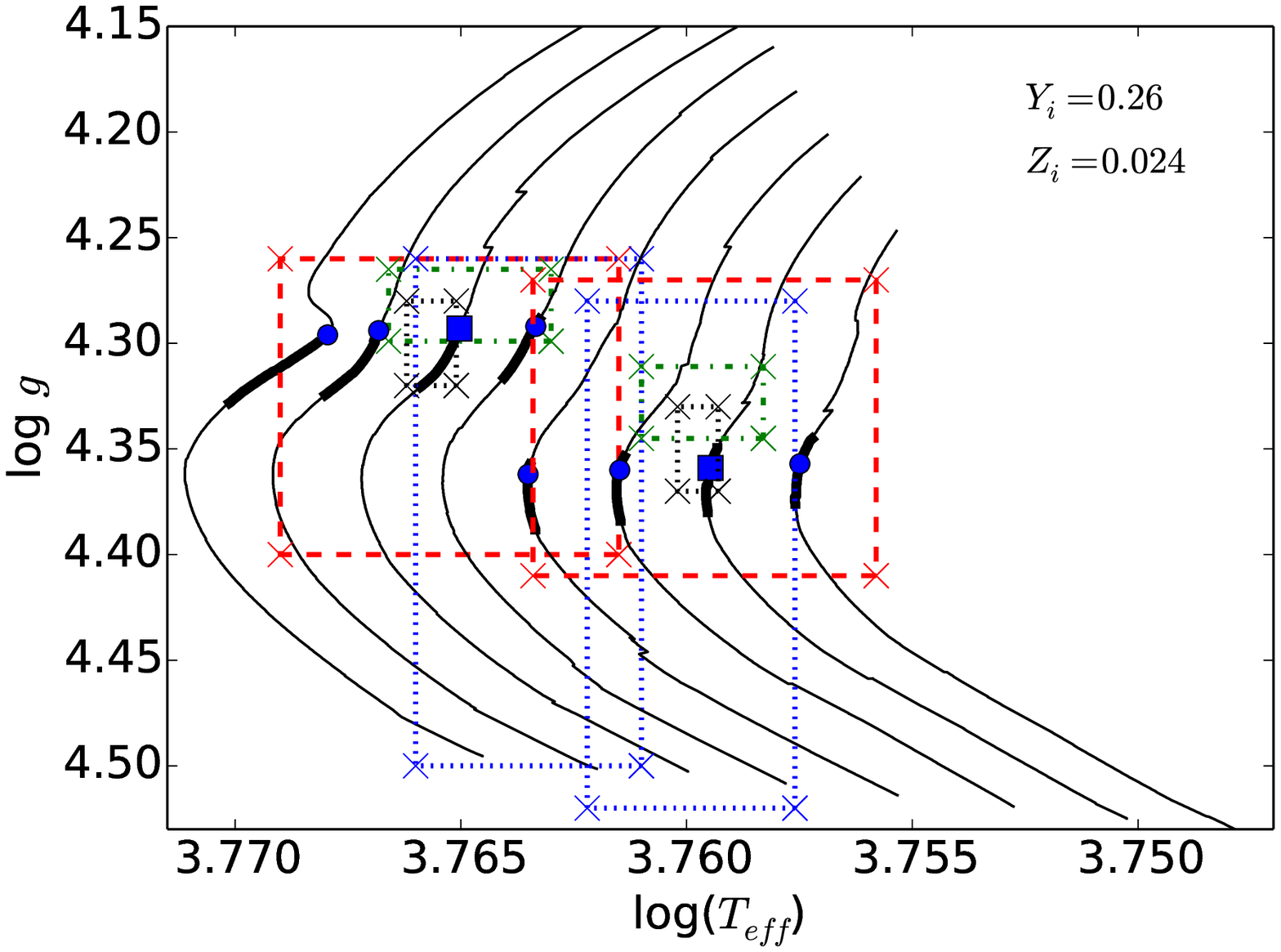}
\caption{Evolutionary tracks for models from 1.05 to 1.14 $M_{\odot}$ (from right to left) with $Z_i=0.024$ and $Y_i=0.25$ (left panel) and evolutionary tracks for models from 1.05 to 1.12 $M_{\odot}$ (from right to left) with $Z_i=0.024$ and $Y_i=0.26$ (right panel) . The error boxes are those of \cite{ramirez09} (red dashed lines), \cite{ramirez11} (green dot-dashed lines), \cite{schuler11} (blue dotted lines) and \cite{tuccimaia14} (black dotted lines). The blue dots indicate models with the right large separation, taking \cite{kjeldsen08} corrections into account. The blue squares correspond to models that also have the right small separations and best fit the Echelle diagram. The black thick segments of each line indicate the models whose radii are consistent with the interferometric determinations of \cite{white13}.}
\label{trace}
\end{center}
\end{figure*}

\begin{figure*}
\begin{center}
\includegraphics[width=0.49\textwidth]{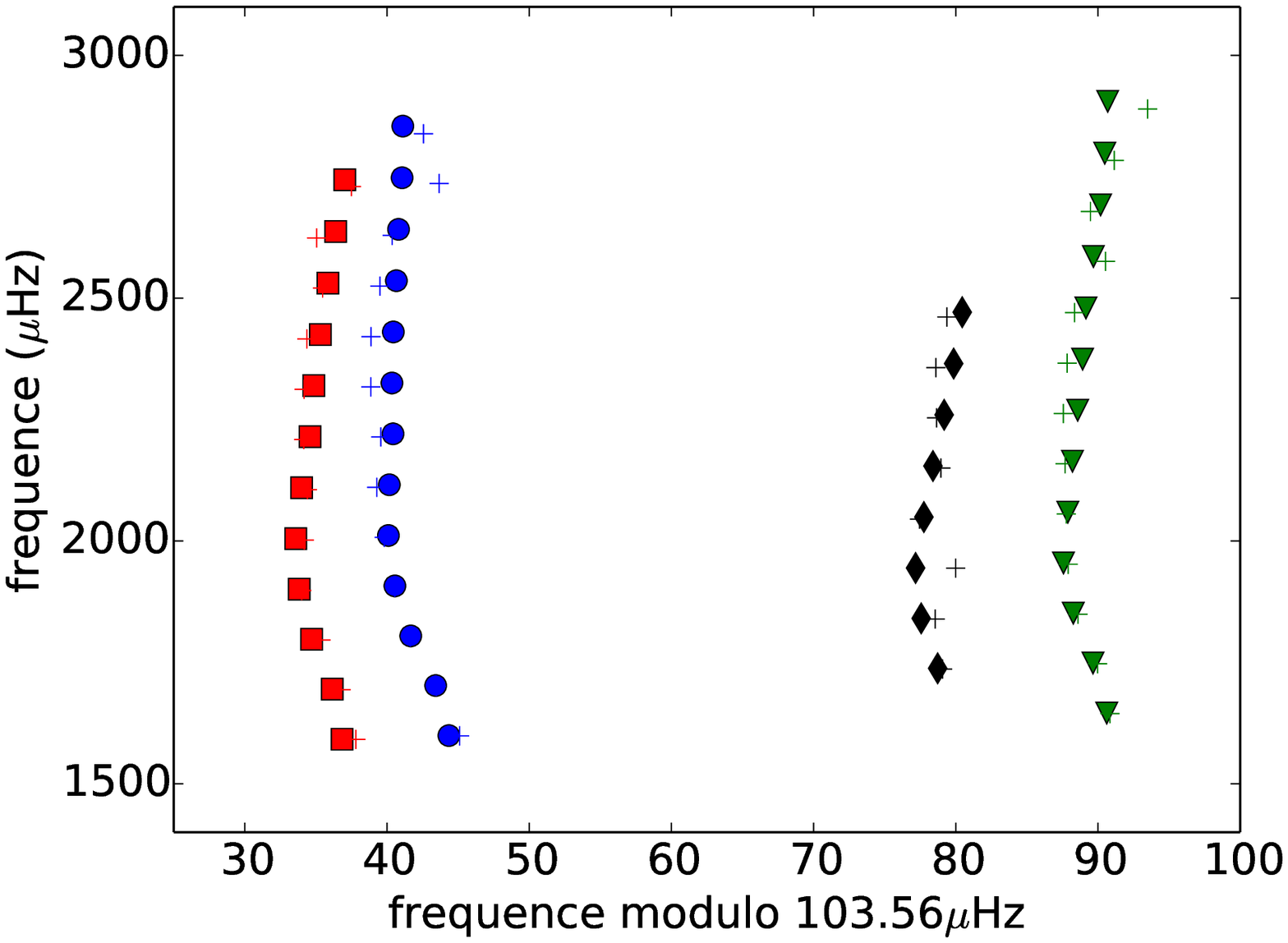}
\includegraphics[width=0.49\textwidth]{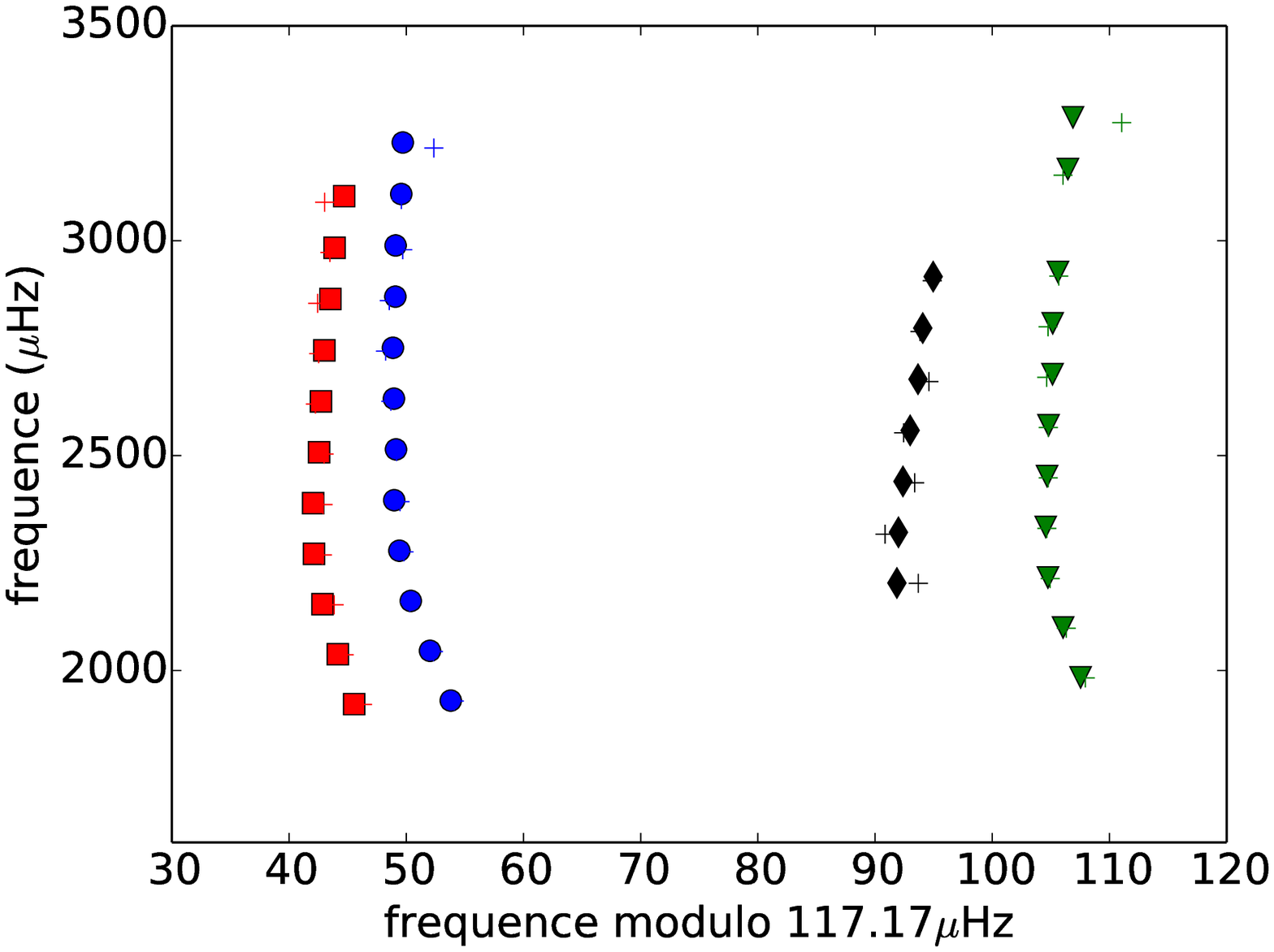}
\caption{Echelle diagrams for 16 Cygni A (left panel) and 16 Cygni B (right panel). The observed frequencies are represented by crosses.The frequencies computed for the models with Y=0.26 and Z=0.024 are represented by blue dots (l=0), green triangles (l=1), red squares (l=2), and black diamonds (l=3).}
\label{ED}
\end{center}
\end{figure*}

The stellar oscillation modes were derived for each model using the PULSE code \citep{brassard08}. The frequencies were corrected for surface effects in the way proposed by \citet{kjeldsen08}.

We computed models with masses ranging from 1.05 to 1.14$M_{\odot}$. The initial helium mass fraction $Y_i$ was varied from 0.245 to 0.26 and the heavy element mass fraction $Z_i$ from 0.023 to 0.025 (initial mass fraction of all elements heavier than helium). 

Evolutionary tracks computed for two different initial compositions $Y_i$=0.25; $Z_i$=0.024 and $Y_i$=0.26; $Z_i$=0.024 are presented in Fig. \ref{trace}. 

The observational uncertainties on the oscillation frequencies for $l=0$ to $l=2$ modes lie between 0.1 to 1.45 $\mu$Hz \citep{metcalfe12}. The derived large separations $\Delta \nu$ for 16 Cygni A and for 16 Cygni B are $\Delta \nu_{obs}=103.56\pm0.10~\mu Hz$ and $\Delta \nu_{obs}=117.17\pm0.10~\mu Hz$, respectively. The models with large separations that fit these values are represented by blue dots in Fig. \ref{trace}. They include the corrections for surface effects as proposed by \citet{kjeldsen08}. Neglecting this effect would lead to models slightly above the blue dots, with a small difference in age of 150 million years on average.
We note that all the models presented in Fig. \ref{trace} lie on horizontal lines, which means that they all have about the same surface gravity. This is a well-known result of asteroseismology. The asymptotic treatment of the oscillations (e.g. \citealt{tassoul80}) shows that the large separation $\Delta \nu$ directly gives the average stellar density. In the range of our possible models, the variations in radii are small so that log $g$ is also nearly constant. In other words, if any parameter of the model is modified, for instance, the initial Y value, the other parameters such
as age adjust to obtain a model with the same large separation, and thus the same gravity. Spectroscopy is only used in our description to constrain the effective temperatures.

 The uncertainty on the log $g$ values derived from the observed seismic frequencies for each track is about $10^{-2}$. This is the uncertainty on the position of the models represented by blue dots in Fig. \ref{trace}. The corresponding error bars would lie inside the printed symbols. We note that the seismic log $g$ values lie outside the ranges given by \cite{ramirez11} (see Fig. \ref{trace}), which suggests that their uncertainties are underestimated. 

We then derived the best of all these models for both stars. We first did it independently for each $Y_i$ value. The best models (represented by squares) were selected first because their small separations $\delta\nu_{0,2}$ best fit the observed ones, second because their echelle diagrams best fit the observed ones, according to $\chi^2$ minimisations performed between the observed and the modelled frequencies.

When compared with spectroscopic observations (see Fig. \ref{trace}), it is clear that the best models for $Y_i$=0.25  lie inside the
box
reported by \cite{schuler11} for 16 Cygni A alone, not for 16 Cygni B, and that they are outside the box derived by \cite{tuccimaia14} for both stars. In contrast,  for $Y_i$=0.26  lie inside all the boxes for both stars. For this reason, we find that the $Y_i$=0.26 value is more probable than that of $Y_i$=0.25.

The echelle diagrams corresponding to the best case computed with $Y_i$=0.26 are presented in Fig. \ref{ED} for 16 Cygni A (left panel) and 16 Cygni B (right panel). The fits between the computed and observed frequencies are very good for both stars. The corresponding stellar masses are $M_A=1.10M_{\odot}$ and $M_B=1.06M_{\odot}$.

These values and the other parameters obtained for these best models are given in Table \ref{table:2}. The uncertainties we
list in this table correspond to the possible range of parameters of all the computed models, which have large and small separations in the observational uncertainties.

We must note that the surface helium abundances derived from these computations are slightly subsolar, while the spectroscopic parameters given in the literature have been computed using a solar helium value. It would be interesting in the future to iterate with spectroscopists and see how such a helium difference could influence the derived effective temperature range.

\begin{table}
\caption{Properties of 16 Cygni A and B from this work} 
\label{table:2}
\centering
\begin{tabular}{c c c}
\hline  
 & 16 Cygni A & 16 Cygni B   \\
\hline
\smallskip 
$T_{\rm eff}$(K) & $5821\pm25$ & $5747\pm25$  \\
\smallskip       
log $g$ & $4.29\pm0.01$ & $4.36\pm0.01$  \\       
\smallskip
Mass ($M_{\odot}$) & $1.10\pm0.01$ & $1.06\pm0.01$  \\        
\smallskip
Radius ($R_{\odot}$) & $1.24\pm0.01$ & $1.13\pm0.01$  \\        
\smallskip
Luminosity ($L_{\odot}$) & $1.58\pm0.03$ & $1.25\pm0.03$  \\        
\smallskip
Age (Gyr) & $6.4\pm0.4$ & $6.4\pm0.4$  \\     
\smallskip
$Z_{i}$ & $0.024$ & $0.024$  \\        
\smallskip
$Y_{i}$ & $0.26$ & $0.26$  \\  
\smallskip
$Z_{surf}$\tablefootmark{a} & $0.0221$ & $0.0223$  \\        
\smallskip
$Y_{surf}$\tablefootmark{a} & $0.2226$ & $0.2265$  \\       
\hline
\end{tabular}
\tablefoot{\tablefoottext{a}{Values at the age of best models}
}
\end{table}

\section{Heavy elements and lithium abundances in 16 Cygni A and B, observational and theoretical discussions}

\subsection{Spectroscopic observations}

The abundances of heavy elements in 16 Cygni A and B have been a subject of debate for 15 years. \cite{deliyannis00} reported an iron overabundance higher in 16 Cygni B than in 16 Cygni A, which has not been confirmed by more recent papers.
Whereas \citet{schuler11} did not report any difference between 16 Cygni A and B for the abundances of heavy elements, \citet{ramirez11} and \cite{tuccimaia14} claimed that the heavy elements are overabundant in 16 Cygni A compared to 16 Cygni B. 
In any case, all observers agree that lithium is more depleted in 16 Cygni B than in 16 Cygni A by a large factor. Lithium has not been detected in B, whereas its abundance in A is slightly higher than that of the Sun. This difference between these very similar stars is difficult to account for using traditional explanations of lithium depletion in G stars.

In the following, we discuss the influence of atomic diffusion and mixing on the lithium and heavy elements abundances, and we show how the lithium difference between the two stars may be explained. We also discuss the consequences of these effects for heavy elements.

\subsection{Influence of atomic diffusion on the abundances of heavy elements and lithium}

\begin{figure}
\begin{center}
\includegraphics[width=0.5\textwidth]{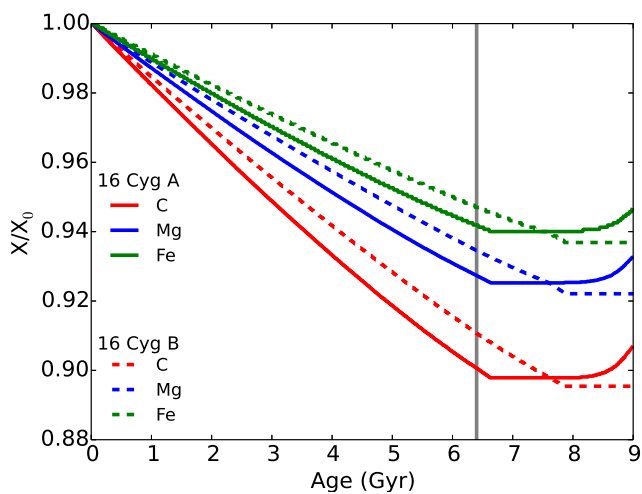}
\caption{Abundance evolution of C, Mg, and Fe in our best models of 16 Cygni A and B, under the influence of atomic diffusion below the convective zone, when no mixing is taken into account.}
\label{xx0}
\end{center}
\end{figure}

Our models include the computation of detailed atomic diffusion for a large number of elements, as discussed in Sect. 3.1. 
Figure \ref{xx0} presents the surface abundance evolution for three of these elements, C, Mg, and Fe, in both 16 Cygni A and B. As the mass of 16 Cygni A is higher than that of 16 Cygni B, the convective zone is smaller and the diffusion processes faster. As a consequence, the surface abundances predicted by the models are higher in 16 Cygni B than in 16 Cygni A by $\sim0.002$ dex at the age of our best models (vertical line). Later on in the evolution, at around 8 Gyr, the situation is reversed because the surface convective zone sinks more quickly in the more massive star. This occurs in models that are too old to account for the observations, however.

\begin{figure}
\begin{center}
\includegraphics[width=0.5\textwidth]{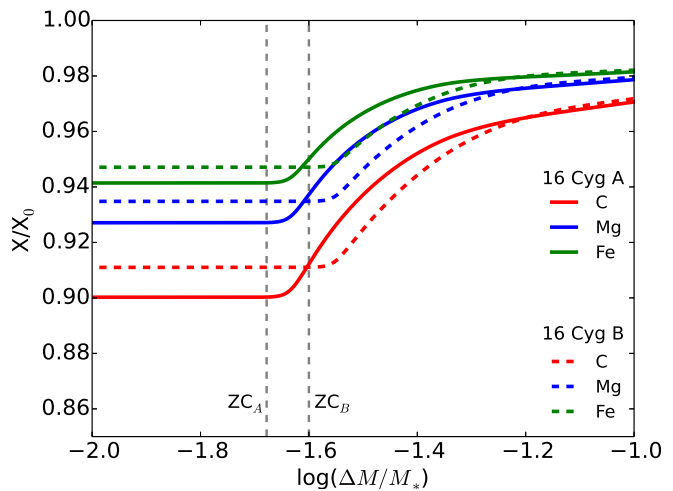}
\caption{Abundance profiles C, Mg, and Fe inside the stars under the same conditions as for Fig. \ref{xx0}. }
\label{abprofil}
\end{center}
\end{figure}

\begin{figure}
\begin{center}
\includegraphics[width=0.5\textwidth]{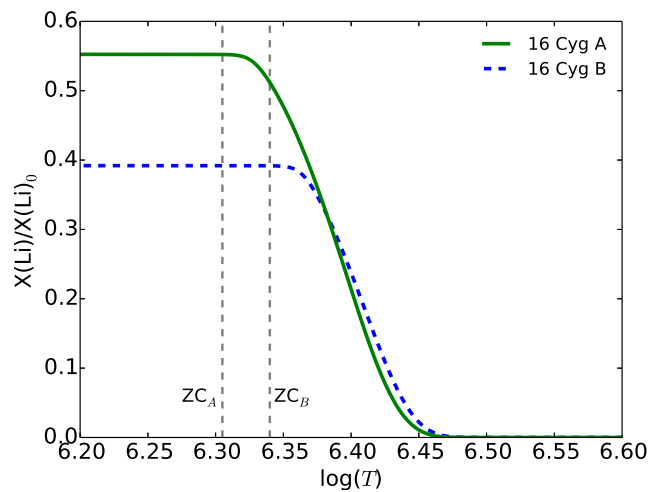}
\caption{Abundance profiles of lithium under the same conditions as for Fig. \ref{xx0}.}
\label{LiLi0}
\end{center}
\end{figure}

Figure \ref{abprofil} displays the abundance profiles for the same three elements C, Mg, and Fe in the best models for 16 Cygni A and B. The bottoms of the convective zones are shown as vertical dashed lines. The difference of heavy elements abundances is due to the difference in surface convective zone depth, which is deeper in 16 Cygni B than A (grey dashed line).
Atomic diffusion coupled with nuclear destruction also has an effect on the abundance of $^7$Li (see Fig. \ref{LiLi0}). In this case, the final abundance of $^7$Li is lower in 16 Cygni B than in 16 Cygni A. This is not enough to account for the observations. First lithium is depleted in 16 Cygni A by a factor smaller than two, while in the real star it is depleted by 100. Second the depletion ratio between the two star is only $\sim1.4,$ while the observations show a ratio of at least 4.7. Extra mixing below the convective zones is clearly needed to account for the observations.

\subsection{Mixing processes below the outer convective zone}

Many papers have been published in the past 30 years on the subject of the mixing processes that may occur below the convective zones of solar type stars and result in lithium destruction in their outer layers. The most common one is rotation-induced mixing (e.g.\citealt{vauclair88,pinsonneault90,charbonnel92,charbonnel94,castro09}).

The rotation periods of 16 Cygni A and B have recently been measured with high precision using asteroseismology \citep{davies15}. They are very similar, 23.8 days for A, 23.2 days for B, with a difference smaller than the uncertainties. The authors have also determined their inclination angles, 56 degrees for A, 36 degree for B. Using the interferometric radius, the resulting $v~\rm sin$ $i$ is 2.23 km.s$^{-1}$ for A and 1.27 km.s$^{-1}$ for B.

When studying rotation-induced mixing, it appears that the efficiency of the mixing is related to the local linear rotation velocity in such a way that it increases with radius. The bottom of the convective zone lies at a larger radius in A than in B, but the mixing is also more efficient. Simple expressions of the mixing diffusion coefficients (e.g. \citealt{zahn92}) include the factor ($\Omega^2 R^3/GM$), which is about 1.3 larger for 16 Cygni A than for 16 Cygni B. As already discussed by \cite{deliyannis00}, it does not seem possible to account for the lithium abundance ratio between A and B by such a simple process. It would not be realistic to invoke a stronger rotation-induced mixing effect in B than in A simply to account for the observations. Another process is clearly needed. 
We do not intend to discuss all these processes here. We only simulate turbulent mixing by using a turbulent diffusion coefficient adjusted to obtain a lithium destruction by a factor 100 in 16 Cygni A, and we used the same to deduce the lithium destruction in 16 Cygni B. As in the simulations of \cite{richer00}, we used a simple form $D_T=\omega D\rm (H_e)$ $\left(\frac{\rho_0}{\rho}\right) ^{n}$ with n = 3, $\omega=325$, $D$(He) the helium diffusion coefficient and $\rho_0$ the density at the bottom of the outer convective zone.

\begin{figure}
\begin{center}
\includegraphics[width=0.5\textwidth]{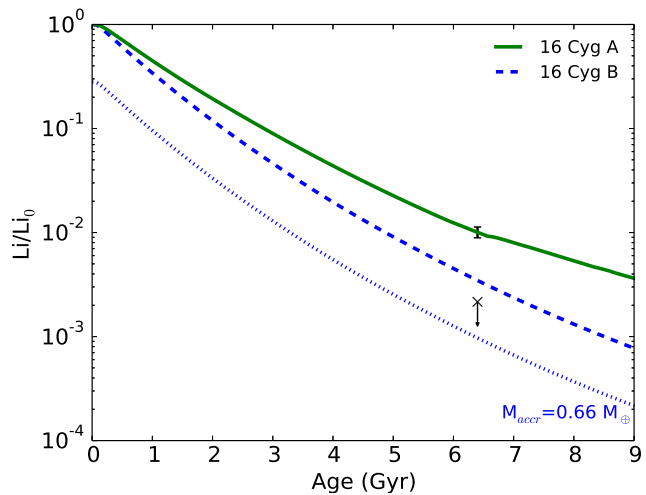}
\caption{Evolution of lithium surface abundances including a mixing at the bottom of the outer convective zone needed to reproduce
the 16 Cygni A abundance in the two models (green solid line and blue dashed line) (Sect. 4.3). Evolution of lithium surface abundances including accretion of 0.66$~M_{\oplus}$ for 16 Cygni B (blue dotted line) at the beginning of the main sequence, but with the same mixing as before (Sect. 5). Abundances determined using observations from \cite{king97} are represented by black crosses.}
\label{Lit}
\end{center}
\end{figure}

Figure \ref{Lit} displays the resulting surface lithium abundance evolution in the two stars (green solid line and blue dashed line). As expected, lithium is more destroyed in 16 Cygni B than in 16 Cygni A, but only by a factor 2.9, not enough to account for the observations of an abundance ratio higher than 4.7 at the age of the stars. 

\begin{figure}
\begin{center}
\includegraphics[width=0.5\textwidth]{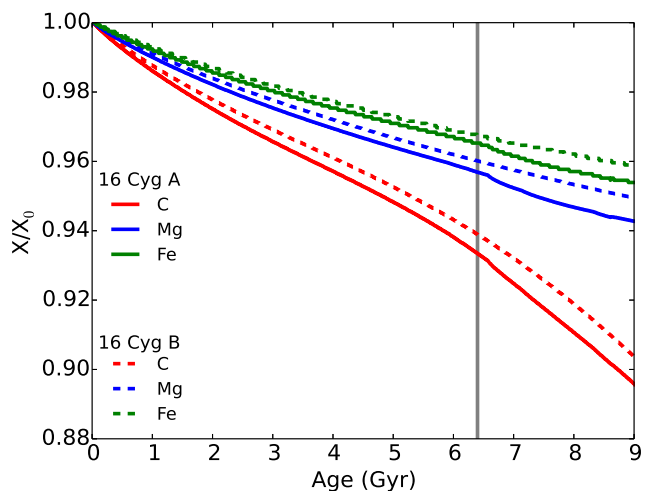}
\caption{Abundance evolution of C, Mg, and Fe in our best models of 16 Cygni A and B, under the influence of atomic diffusion below the convective zone, when mixing is taken into account. }
\label{xt}
\end{center}
\end{figure}

Meanwhile, the abundance of heavy elements is also modified, but the abundances in 16 Cygni B may not become lower than in 16 Cygni A (see Fig. \ref{xt}).

\section{Fingering convection and lithium destruction induced by planetary accretion}

We now return to the fact that the star 16 Cygni B hosts a giant planet while 16 Cygni A does not. As discussed in the introduction, the red dwarf that orbits 16 Cygni A may be the reason for the fact that the main star could not develop any planetary disk.

In this situation, it is quite possible that in its early period on the main sequence, 16 Cygni B was able to accrete some matter from its planetary disk, which could not occur for 16 Cygni A.
As discussed by \citep{theado12bis}, the accretion of heavy planetary matter onto a stellar surface builds an unstable compositional gradient at the bottom of the surface convective zone that triggers fingering (thermohaline) convection. This mixing leads to complete dilution of the heavy matter inside the star, so that no signature appears at the surface for the heavy elements. On the other hand, it may lead to extra lithium depletion because the mixing continues
down to the destruction layers.

\subsection{Computations of fingering convection}  

Fingering convection may occur in stars every time a local accumulation of heavy elements appears in the presence of a stable temperature gradient. It occurs in particular in the case of the accretion of planetary matter \citep{vauclair04, garaud11, deal13}. 

Fingering convection is characterised by the so-called density ratio $R_0$ , which is the ratio between the thermal and $\mu$-gradients: 
\[
R_0=\frac{\nabla - \nabla_{ad}}{\nabla_{\mu}}. 
\] 
The instability can only develop if this ratio is higher than one and lower than the Lewis number, which is the ratio of the thermal to the molecular diffusivities. In this case, a heavy blob of fluid falls down inside the star and continues to fall because it exchanges heat more quickly than particles with the surroundings. If $R_0$ is smaller than one, the region is dynamically convective (Ledoux criterium), and if it is larger than the Lewis number, the region is stable.

Various analytical treatments of fingering convection in stars were given in the past, leading to mixing coefficients that differed by orders of magnitude \citep{ulrich72,kippenhahn80}. More recently, 2D and 3D numerical simulations were performed, converging on coefficients of similar orders \citep{denissenkov10,traxler11}. 

Here we used the recent prescription given by \citet{brown13}, which has been confirmed by the 3D simulations of \citealt{zemskova14}.

\subsection{Effect of accretion-induced fingering convection on the lithium abundance of 16 Cygni B} 

We computed stellar models of 16 Cygni B with the assumption of accretion of planetary matter at the beginning of the main-sequence phase. These models included the treatment of fingering convection in the case of an inversion of the mean molecular weight gradient. We tested different accretion masses to characterise their effect on the lithium surface abundance due to fingering convection. In these computations, the accreted matter was assumed to have an Earth-like chemical composition \citep{allegre95}. Changing the relative abundances of the accreted heavy elements has a very weak effect because their contribution to the mean molecular weight may be slightly different. For example, decreasing the iron abundance by a factor two compared to other elements modifies the computed accretion mass by less than $10\%$.

\begin{figure}[h]
\begin{center}
\includegraphics[width=0.5\textwidth]{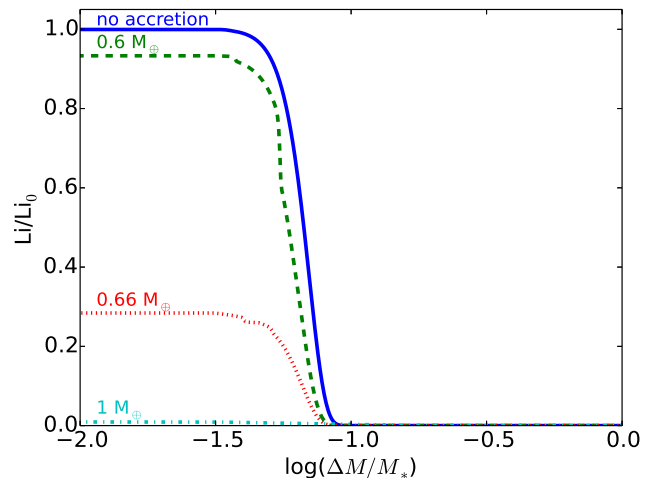}
\caption{Lithium abundance profiles after the accretion of different masses in the model of 16 Cygni B: no accretion (blue solid line), 0.6$~M_{\oplus}$ (green dashed line), 0.66$~M_{\oplus}$ (red dotted line), and  1$~M_{\oplus}$ (cyan dotted-dashed line) at the beginning of the MS.}
\label{Li-accr}
\end{center}
\end{figure}

Abundance profiles of lithium after the accretion of various masses of planetary matter are presented in Fig. \ref{Li-accr}.

An accreted mass lower than or equal to 0.6$~M_{\oplus}$ does not have any real effect on the lithium abundance because it reduces its surface value by a factor lower than 10 percent. The mixing is not efficient enough to mix the lithium down to the destruction layers. For masses higher than 0.6$~M_{\oplus}$, the lithium destruction becomes important and reaches a factor 3.5 for an accreted mass of 0.66$~M_{\oplus}$ and more than a factor 100 for an accreted mass of 1$~M_{\oplus}$.

Thus a small amount of planetary matter is enough to significantly
reduce the lithium surface abundance in this type of stars. 

After the accretion episode in 16 Cygni B, the lithium abundance continues to decrease due to other extra mixing processes in the two stars, as discussed in the previous section. We plot in Fig. \ref{Lit} the lithium surface abundance evolution with the assumption of an accretion episode of 0.66 $M_{\oplus}$ (blue dotted line), which is enough to account for the observations. As the lithium observation in 16 Cygni B is only an upper limit, any accretion mass higher than this value can explain the abundance difference in the two stars. 

\subsection{Beryllium }

Beryllium has been detected in 16 Cygni A and B by \cite{deliyannis00}, who found that the difference between the two abundances, if any, must be smaller than 0.2 dex.  
Beryllium is destroyed by nuclear reactions at a temperature of $T\sim3.5~10^6 K$, which is higher than the temperature needed
to destroy lithium. The fact that it is not depleted gives an upper limit on the accreted mass that must not lead to mixing down to its destruction region.

\begin{figure}[h]
\begin{center}
\includegraphics[width=0.5\textwidth]{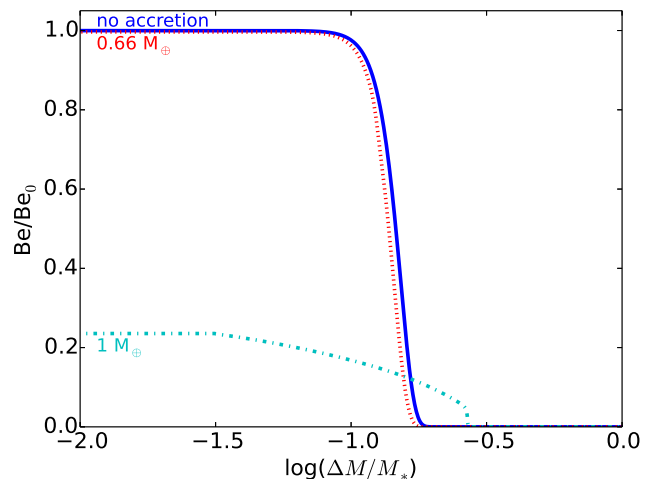}
\caption{Beryllium abundance profiles after the accretion of different masses: no accretion (blue solid line), 0.66$~M_{\oplus}$ (red dotted line), and  1$~M_{\oplus}$ (cyan dot-dashed line) at the beginning of the MS.}
\label{Be-accr}
\end{center}
\end{figure}

To characterise this effect, we plot the same figure as Fig. \ref{Li-accr} for beryllium (see Fig. \ref{Be-accr}). An important result is that for an accretion of 0.66$~M_{\oplus}$ (red dotted line) beryllium is not destroyed by nuclear reactions because fingering convection does not mix the stellar matter deep enough. However, for an accretion of 1$~M_{\oplus}$ (cyan dot-dashed line) beryllium is already reduced by a factor 5. If the beryllium abundance determination of the two components of the 16 Cygni system is confirmed, it may lead to a precise determination of the accreted mass.

\section{Summary and discussion}

The 16 Cygni system is particularly interesting for comparative studies of stars with and without planets. 
The two main stars of this system have been observed in several ways, leading to very precise constraints. 

The most striking feature is the lithium abundance, which is slightly higher than solar for 16 Cygni A and lower by a factor of at least 4.7 in 16 Cygni B (\citealt{friel93,king97}). The lithium value given for this later star is an upper limit, which
means that it could be completely lithium depleted.

We here presented detailed computations of these two stars, leading to models that precisely fit their asteroseismic frequency determinations, as well as their radii, luminosities, effective temperatures, and gravities. These models were computed by fully taking into account atomic diffusion of helium and heavy elements. Their characteristics are given in Table \ref{table:2} \footnote{\textbf{Note added in proof:}
After this paper was accepted, Travis Metcalfe draw our attention to his new article accepted for publication in ApJ  "Asteroseismic modeling of 16 Cyg A \& B using the complete Kepler data set" by T.S. Metcalfe, O.L. Creevey \& G.R. Davies, arXiv:1508.00946v2. The authors revise the data given in \cite{metcalfe12}, particularly the age ($7.0 \pm 0.3 $ Gyr) and the composition ($Z = 0.021\pm0.002$, $Yi = 0.25\pm0.01$). They acknowledge that the helium content they derive may be slightly underestimated due to their neglecting atomic diffusion of heavy elements in the models. It is interesting to stress that we indeed find a slightly higher helium content and a younger age than they.}. 

Then we discussed the importance of the lithium observations for the two stars. We first showed that the very large lithium depletion observed in 16 Cygni B cannot be accounted for by the classical means of rotation-induced mixing or similar types of extra mixing. The observable parameter difference between them is too small to explain such a large difference in lithium as observed. We suggest that the accretion of heavy matter onto the planet host star 16 Cygni B in its early main-sequence phase induced a special kind of mixing, namely fingering convection, which led to a large lithium destruction on a relatively small timescale (cf. \citealt{vauclair04,theado12bis}).

We recall that when a star accretes heavy matter onto lighter one, the induced $\mu$-gradient leads to a specific kind of hydrodynamical instability, called fingering convection or thermohaline convection, which mixes all the accreted matter downwards. Contrary to what is often assumed, the accreted heavy elements do not leave any signature in the stellar outer layers because of this mixing. On the other hand, the same mixing may transport lithium down to the layers where it is destroyed by nuclear reactions. 

We suggest that during the evolution of the planetary disk, at the beginning of the main-sequence phase, 16 Cygni B may have swallowed a fraction of an Earth-like planet that has led to fingering convection below the convective zone. This short-time efficient extra-mixing allowed the transport of lithium down to the nuclear destruction region. We have seen that two thirds of an Earth-mass planet would be enough to account for the observed upper value of lithium in 16 Cyg B, as shown in Fig. \ref{Lit}. 
Observations of beryllium have been reported in these two stars by \cite{deliyannis00}. They found no difference in beryllium
between the two stars. If confirmed, this means that the accretion of planetary matter could not have exceeded one Earth mass. 

We also computed the detailed variations of the heavy element abundances due to atomic diffusion along the two evolutionary tracks leading to the best models for 16 Cygni A and B. Atomic diffusion alone leads to carbon depletion by about 0.05 dex and to magnesium depletion by about 0.03 dex. The resulting abundances in the two stars are very similar, as observed by \cite{schuler11}. Differences as suggested by \citealt{ramirez11} and \citealt{tuccimaia14} cannot be explained in this context. 

In summary, the 16 Cygni system gives evidence of fingering convection induced by the accretion of planetary matter, which is probably the unknown process invoked in the past by several authors to account for the large lithium differences between the two stars.
 
This result can be generalized to all planet-hosting stars, which may accrete planetary matter in a random way, with various accretion rates. As suggested by \citet{theado12}, this could lead to an average lithium abundance that is lower in planet-hosting stars than in other stars, as claimed by \citet{israelian09} and \citet{delgado14}. In this framework, precise lithium abundances in seismically observed stars with or without observed planets would be very useful. Furthermore, beryllium observations in the same stars would help constraining the mass accreted onto the star.

\begin{acknowledgements}
We thank the "Programme National de Physique Stellaire" (PNPS) of CNRS/INSU (France) for financial support. We also warmly thank Piercarlo Bonifacio for very useful comments on the first version of this paper.
\end{acknowledgements}  
  
\bibliographystyle{aa} 
\bibliography{biblio.bib} 

\end{document}